\documentstyle[emulateapj,psfig]{article}

\def\etal{et al.}

\def\chandra{{\sl Chandra}}
\def\Chandra{{\sl Chandra}}
\def\gsim{\lower 2pt \hbox{$\, \buildrel {\scriptstyle >}\over
{\scriptstyle \sim}\,$}}
\def\lsim{\lower 2pt \hbox{$\, \buildrel {\scriptstyle <}\over
{\scriptstyle \sim}\,$}}

\newcommand{\casa}{\hbox{Cas~A}}

\slugcomment{To appear in the Astrophysics Journal Letters}

\lefthead{Gotthelf \etal}
\righthead{Forward and Reverse Shocks in \casa}

\begin{document}

\title{Chandra Detection of the Forward and Reverse Shocks in Cassiopeia~A }

\author{E. V. Gotthelf$^1$, B. Koralesky$^2$, L. Rudnick$^2$,  T.W. Jones$^2$, U. Hwang$^{3,4}$, R. Petre$^3$}
\altaffiltext{1}{Columbia Astrophysics Laboratory, Columbia University, 550 West 120$^{th}$ Street, New York, NY \ 10027, USA; evg@astro.columbia.edu} 
\altaffiltext{2}{Department of Astronomy, University of Minnesota, 116 Church Street SE, Minneapolis, MN \ 55455}
\altaffiltext{3}{Laboratory for High Energy Astrophysics, Goddard Space Flight Center, Greenbelt, MD \ 2077}
\altaffiltext{4}{Department of Astronomy, University of Maryland, College Park, MD \ 20742}

\begin{abstract} 

We report the localization of the forward and reversed shock fronts in
the young supernova remnant \casa\ using X-ray data obtained with the
\Chandra\ Observatory.  High resolution X-ray maps resolve a
previously unseen X-ray feature encompassing the extremity of the
remnant.  This feature consists of thin, tangential wisps of emission
bordering the outer edge of the thermal X-ray and radio remnant,
forming a circular rim, $\sim 2\farcm7$ in radius.  Radio images show
a sharp rise in brightness at this X-ray rim, along with a large jump
in the synchrotron polarization angle. These characteristics suggest
that these wisps are the previously unresolved signature of the
forward, or outer, shock. Similarly, we identify the sharp rise in
emissivity of the bright shell for both the radio and X-ray line
emission associated with the reverse shock. The derived ratio of the
averaged forward and reverse shock radii of $\sim$~3:2 constrains the
remnant to have swept up roughly the same amount of mass as was
ejected; this suggests that \casa\ is just entering the Sedov phase.
Comparison of the X-ray spectra from the two shock regions shows that
the equivalent widths of prominent emission lines are significantly
lower exterior to the bright shell, as expected if they are
respectively identified with the shocked circumstellar material and
shocked ejecta.  Furthermore, the spectrum of the outer rim itself is
dominated by power-law emission, likely the counterpart of the
non-thermal component previously seen at energies above $\sim$ 10 keV.

\end{abstract}


\section {Introduction}

The young supernova remnant Cassiopeia~A (\casa) is thought to result
from the destruction of a massive star whose explosion was likely
recorded by the Dutch astronomer Flamsteed in 1680 (Ashworth 1980;
Fabian \etal\ 1980).  As one of the brightest radio and X-ray emitters
in the sky, \casa\ is perhaps the best studied shell-like supernova
remnant (SNR). Located along the Galactic plane at a distance of $\sim
3.4$ kpc (Reed \etal\ 1995), \casa\ appears today as a nearly
circular, $3^{\prime}$ diameter bright ring, with a low surface
brightness $5^{\prime}$ diameter radio and X-ray plateau.
Multi-wavelength analysis has fleshed out a picture of an expanding
shell of supernova ejecta, whose X-ray emission is dominated by a
thermal spectrum characteristic of a shocked plasma, rich in emission
lines from highly ionized atoms (Becker \etal\ 1979); Fesen, Becker \&
Blair 1987).  The radio emission is consistent with synchrotron
radiation from relativistic particles accelerated at shocks or other
structures during the remnant's expansion (Anderson \& Rudnick 1996).

Studies of SNR shock structures can provide important dynamical and
radiative information on the evolutionary state of the remnant (Fabian
\etal\ 1980; Fesen, Becker \& Blair 1987; Braun 1987; Greidanus \&
Strom 1991; Reed \etal\ 1995).  In the standard picture of young SNRs,
we expect to see an outer shock, the blast wave from the explosion
moving into the circumstellar medium, and a reverse shock, responsible
for decelerating and compressing the outflowing ejecta.  Analytical
and numerical work suggests that these two shocks are still generated
even when the ejecta are very inhomogeneous, as in \casa\ (see {\it
e.g.}, Hamilton 1985; Jun, Jones \& Norman 1996).  In this
inhomogeneous environment, the shock structures may become locally
complex, and the ``contact discontinuity'' between shocked ejecta and
circumstellar material becomes a broad, non-uniform region.

In this Letter, we report the detection of sharp, well defined X-ray
emission features encircling the outer boundary of \casa\ which we
argue are the signature of the forward shock front. We use new X-ray
and radio data to measurement the location of both the forward and
reverse shocks and compare these to previous work. The remnant is most
likely in a position-dependent dynamical state in transition from free
expansion to the Sedov phase.

\section {Observations}

We begin our analysis of \casa\ using X-ray data obtained during the
``first light'' observation of the \Chandra\ Observatory (Weisskopf
\etal\ 1996). This observation, which imaged \casa\ using the Advanced
CCD Imaging Spectrometer (ACIS), was the basis of a recent study by
Hughes \etal\ (2000) on the distribution of nucleosynthesis products
in the remnant; the reader is referred to this paper for details about
the observation and its data set.  We confirm our results using a
deep 50 ks archival observation of \casa\ obtained on 30 Jan 2000
using the same observational configuration.  The superb imaging
quality of \chandra\ allowed us to produce two narrow band maps of
\casa\ with arcsecond resolution, one restricted to the spectral
region containing the He-like Si transitions near 1.86 keV (including
the weaker H-like transition at 2.006 keV), and the other restricted
to the continuum-dominated $4-6$ keV band. 
The Si image has been continuum subtracted using the high energy map,
normalized by estimating the continuum component underlying the line
emission spectrum.
These images are displayed in Figure 1, along with a recent high
resolution 4410 MHz radio continuum map obtained with the Very Large
Array (for details see Koralesky \etal\ 2001, in preparation). The
radio map was also used to reference the absolute aspect of the
\chandra\ X-ray image.

\section{Forward and reverse shock identification}

The overall morphology of \casa\ as seen in the three images in
Figure~1 is familiar from earlier X-ray and radio studies. The
emission is dominated by the bright ring, $\sim 1\farcm8$ in radius,
beyond which extends a plateau of low surface brightness.  However,
the two new X-ray images shows a clear difference in detail, first
suggested in Holt \etal\ 1994, with many filaments unique to each map,
especially at the very edge of the plateau. Here, in the X-ray
continuum image, \Chandra's arcsecond angular resolution reveals for
the first time a halo of thin, tangential wisps, roughly $
3^{\prime\prime}$ in width, which encompasses the bulk of the thermal
X-ray and radio emission.  These wisps define the external boundary of
the X-ray remnant centered at $23^h 23^m 26\farcs98 \ +58^{\circ} 48^{\prime}
45^s.9$ (J2000) with a mean radius of $153^{\prime\prime}$,
extending $\pm 12^{\prime\prime}$ (see Fig. 2). The fineness of these
features prevented their detection in earlier, lower resolution, X-ray
images. Traces of the wisps can also be seen in the Si map, consistent
with a lower energy component of the X-ray continuum underlying the
line emission.  Below we identify these wisps with \casa 's forward
shock, where the blast wave from the supernova explosion encounters
the circumstellar medium.

We looked at the radio map for evidence of wisps coincident with those
observed in X-rays; no equivalent feature is seen. In Figure 3, we
plotted the radial profiles of \casa\ in the X-ray and radio
wave-bands centered on the above coordinates.  We find a large
increase in the radio surface brightness interior to the wisps leading
to the plateau prior to the bright ring. For clarity, we restricted
our profiles to an azimuthal range of $-5 < 
\theta\ < -60$ degs, as
the offset bright ring emission is somewhat blurred around the remnant
for this center.  Also plotted in this figure is the synchrotron
polarization angle which reveals a significant jump in angle at the rim,
interior to which the inferred magnetic field is predominantly radial.
The combination of this abrupt jump in polarization angle and the
sharp rise in the radio brightness, both coincident with the newly
resolved rim of thin X-ray wisps, strongly suggest that this emission
feature is the signature of the forward shock in \casa.

We now consider evidence for a reverse shock, for which the clearest
indicator would be a sharp rise in X-ray and/or radio emissivity with
increasing radius.  No such feature has been previously identified in
\casa, partially due to the very patchy emissivity unresolved in
X-rays. This non-uniformity reflects inhomogeneities in the ejecta
and/or the circumstellar medium. The quality of the X-ray
image-spectroscopy with \chandra\ now offers the possibility of
revisiting this issue.

To isolate the reverse shock in \casa\ we deproject the observed
line-of-sight integrated brightness into a one-dimensional emissivity
profile as a function of radius.  An earlier attempt at shell
decomposition by Fabian \etal\ 1980 used the ROSAT \casa\ image to
qualitative assigned the shocks to a two-shell structure. The general
method is to assume that the emissivity (at least for a limited range
in azimuth, in this case) can be modeled as a set of thin, uniform
shells. We perform an iterative decomposition of the brightness
profile as a function of radius into one-pixel-wide constant
emissivity shells. The iteration proceeds until the residual
brightness is lowered to 1\% of its original peak value. Here, we
restrict our analysis to the region around the bright ring ( $100 <
\theta < 250$ degs) which appears to be the region most free of
unrelated filamentary features seen in projection against the ring.

The results of this spatial decomposition are shown in Figure 4. We
take the center of the bright ring as a reasonable approximation to
the reverse shock centroid which we estimate at ($23^h 23^m 25\farcs44
\ +58^{\circ} 48^{\prime} 52.^s3 $; J2000). It is clear that both the
radio and Si emissivity profiles for this sector show a sharp rise at
a radius of $95\arcsec \pm 10\arcsec$, which we identify as the
location of the reverse shock.  The mean ratio of the outer shock to
reverse shock radii is then $\sim$ 3:2 with a $14\%$ range
($153^{\prime\prime} \pm 12^{\prime\prime}$:$95^{\prime\prime} \pm
10^{\prime\prime}$). We note that these emissivity profiles are only
reliable for radii $> 75\arcsec$, interior of which the deconvolution
method applied to our data sets begins to diverge.  There is also
marginal evidence for the reverse shock in the emissivity profiles
using the $4-6$ keV images (not shown).

\section {Discussion}

\subsection{\casa's dynamical state}

The location of the two shock fronts in \casa\ allows us to place
important dynamical constraints on the evolutionary state of the
remnant.  A key model parameter for the evolution of a young
core-collapsed SNR is the ratio of the blastwave radius to that of the
reverse shock, $r_b/r_r$. In many models this ratio characterizes the
transition from the initial free-expansion, ``ejecta dominated'' phase
to the simpler Sedov phase (see Truelove \& McKee 1999 for a review).
For a broad range of assumptions about the density structure of the
stellar envelope and the circumstellar material (CSM), $r_b/r_r$
depends primarily on the ratio $\eta$ of swept-up to ejected
mass. Indeed, the observed shock ratio for \casa\ is consistent with
the models of Truelove \& McKee (1999) for which $\eta \approx 1$.  In
their models, the reverse shock generally penetrates through the
stellar envelope and into the core of the ejecta somewhat sooner than
when $\eta = 1$. The close proximity in \casa\ of the blast shock to
the reverse shock suggest that the amount of swept-up material is
limited, and thus the value of $\eta \lsim 1$.

The ratio $r_b/r_r$ also plays a role in more sophisticated
models. Borkowski \etal\ (1996) examined a CSM model based on the
formation of a spherical shell during the final stages of presupernova
evolution. In their model $r_b/r_r$ is also close to 3:2, but the
inner shell boundary, located near $0.9 r_b$, should be much more
prominent than the reverse shock. Some refinement would be necessary
for this model to be compatible with the new data.

The localization of the shock structures presented here allows us to
compare their apparent geometric offset with the kinematic derived
centers. This provides further insight into the well known large-scale
asymmetries in \casa's expansion (Tuffs 1986; Anderson \& Rudnick
1995; Reed \etal\ 1995; Koralesky \etal\ 1998). The reverse shock is
found to be offset to the northwest by $\sim 0\farcm22$, ($\sim 15\%$
of the ring radius) from the center of the forward shock.
Interestingly, the point source (Tananbaum 1999) is located $\sim
0\farcm11$ in the exact opposite direction from a vector between the
forward to the reverse shock. The size of our derived geometric offset is
found to be consistent with the various kinematic centers summarized
by Reed \etal\ 1995.

These results confirm that the explosion and/or the circumstellar
environment of the precursor were significantly asymmetric.  A strong
asymmetry in velocities is seen in both the compact radio features
(Tuffs 1986; Anderson \& Rudnick 1995) as well as more diffuse
emission (Koralesky \etal\ 2001, in preparation).  Reed \etal\ (1995)
conclude that there must be a density gradient of $\sim 5$
front-to-back across the remnant from asymmetries in the line-of-sight
velocities of the fast optical emission-line knots (e.g. Fesen, Becker
\& Blair 1987). Ongoing two-dimensional velocity analysis of the radio
data and future X-ray velocity data should help elucidate the complex
dynamical state of the remnant.

\subsection{The nature of \casa's shock structures}

It is illustrative to compare representative X-ray spectra from
regions associated with the forward and reverse shocks.  These
selected spectra, shown in Figure 5, indicate clear qualitative
differences, most notably in the strength of the prominent (Si and S)
emission line blends. For the spectrum of the wisps, including the
plateau interior to it, taken from a typical region subtending about
20 degs in the north, the Si and S line equivalent widths are 160
and 140 eV, respectively, while other regions of the rim are nearly
devoid of emission lines (also see Hughes \etal\ 2000).  For the ring,
represented here by a region in the northeast, they are 990 and 730
eV.  Elsewhere in the ring, these may be as much as a factor of two
larger or smaller. This difference is expected if the rim emission
arises predominantly from interstellar material heated by the forward
shock, and the ring emission from reverse-shocked ejecta.

The spectrum of the portion of the rim discussed above can be fitted
with a planar shock model (Borkowski, xspec 11.0) with kT =2.0$\pm$0.5
keV, over a range of {\it nt} values from 0 to
10$^{10.8\pm0.2}$~cm$^{-3}$ s, and N$_H$=1.0$\pm$0.1$\times$10$^{22}$
cm$^{-2}$.  An additional hard component is required, however, for a
reasonable fit. We used a power-law, and found a spectral index
similar to that reported by Allen \etal\ (1997) for the RXTE spectrum
above 10 keV. The flux of this power-law component is at least half,
but can be as much as $70\%$, of the total $0.5-10$ keV flux.  It is
possible that this component could be significantly higher in other
portions of the rim.  In contrast, for the bright ring spectrum shown,
a hard component is not required; the $90\%$ confidence upper-limit
for its flux contribution is $20\%$.  We leave a more detailed spectral
analysis of the shock fronts for a future paper.

The high energy tail component required by the spectra fits to the
wisp region raises the possibility of an X-ray synchrotron component
produced by electrons accelerated to TeV energies in the outer shock,
as seen in SN1006 (see Koyama \etal\ 1995; Reynolds 1996). For \casa,
however, this is somewhat problematic since there is no direct radio
counterpart to the wisp emission seen in the $4-6$ keV X-ray energy
band. A non-thermal bremsstrahlung model provides an alternative
explanation for the high energy X-ray tail (Laming 2001; Bleeker
\etal\ 2001) consistent with the lack of radio counterpart.

It is clear from the new data that the reverse shock forms a distinct
inner boundary for both the hot ejecta producing the X-ray emission
and the amplified magnetic fields producing the radio emission. The
radio structure is more problematic since it depends sensitively on
the strength of the magnetic field, in addition to the results of any
local particle acceleration. We do see evidence for the effect of the
outer shock on the magnetic field through the enhanced alignment of
the polarization angle to the shock normal just inside the rim. But
the field is still largely disorganized, since the degree of
polarization remains weak ($\sim 5\%$, Anderson, Keohane \& Rudnick
1995).  It is likely that the radio emissivity is being controlled
mainly by turbulent field amplification between the two shocks.

In summary, the key results of this study are: 1) the discovery of a
thin, bright X-ray wisps which we interpret as the forward shock, 2)
the identification of a sharp rise in radio and X-ray line emissivity
at the inner edge of the bright ring, which we associate with the
reverse shock, 3) the determination of bulk differences in the X-ray
spectra associated with these shocks, 4) and the inference that \casa\
has currently swept up approximately its own ejected mass.
 
\acknowledgements

We gratefully acknowledge the \Chandra\ team for making available the
public data used herein. This work was funded in part by NASA LTSA
grants NAG5-7935 (E.V.G.), NASA GSRP (B.T.K.), and the NSF under grant
AST96-19438 (U Minn.).

%\end{document}

%\vfil\eject

\onecolumn

\begin{figure} 
\centerline{
 \psfig{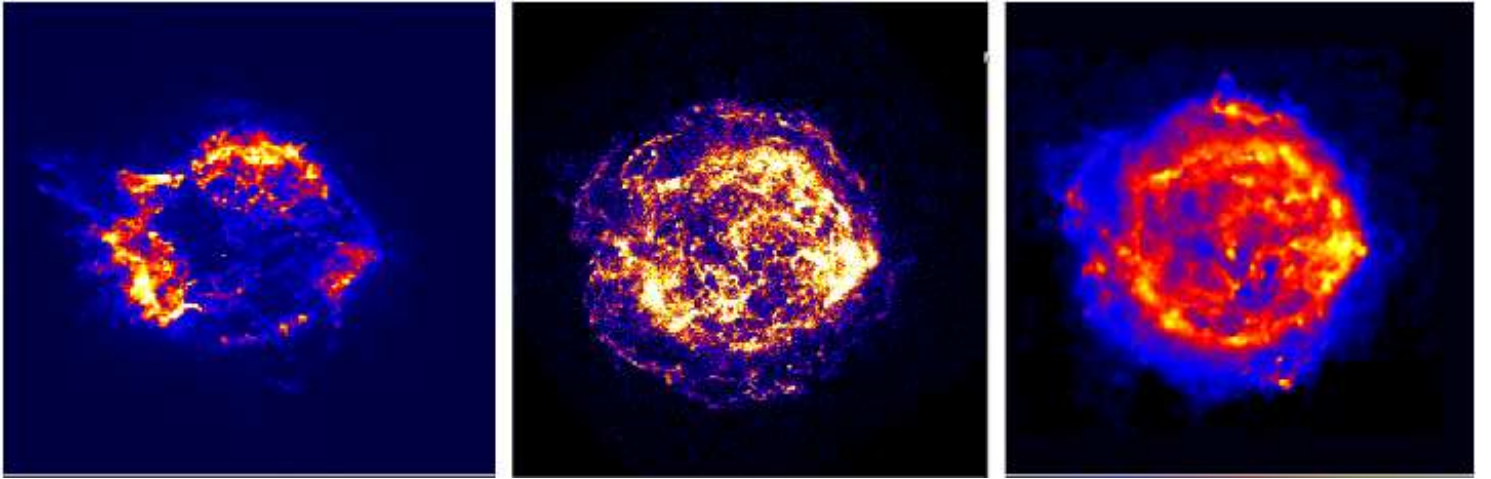} 
}
\caption{\Chandra\ ACIS narrow-band X-ray images and radio map
of \casa.  These images show the identical field and are displayed
with square root intensity scaling with the contrast adjusted to
highlight the low surface brightness regions of the remnants.
{\it Left --} the continuum subtracted Silicon X-ray imiage restricted to the 
bright He-like $K_{\alpha}$ ($1.68$ keV) spectral line feature. 
{\it Middle --} high-energy X-ray continuum
emission map of \casa\ in the line-free spectral region between $4-6$
keV plotted on the same scales as in the left panel. Fine wisps are
evident beyond the main shell-like structure of the supernova
remnant. {\it Right --} Radio map at 4.4 GHz from VLA, epoch 1997, $\approx
1.5\arcsec$ resolution.
\label{fig1}}
\end{figure}

\twocolumn

\begin{figure} 
\centerline{
\psfig{figure=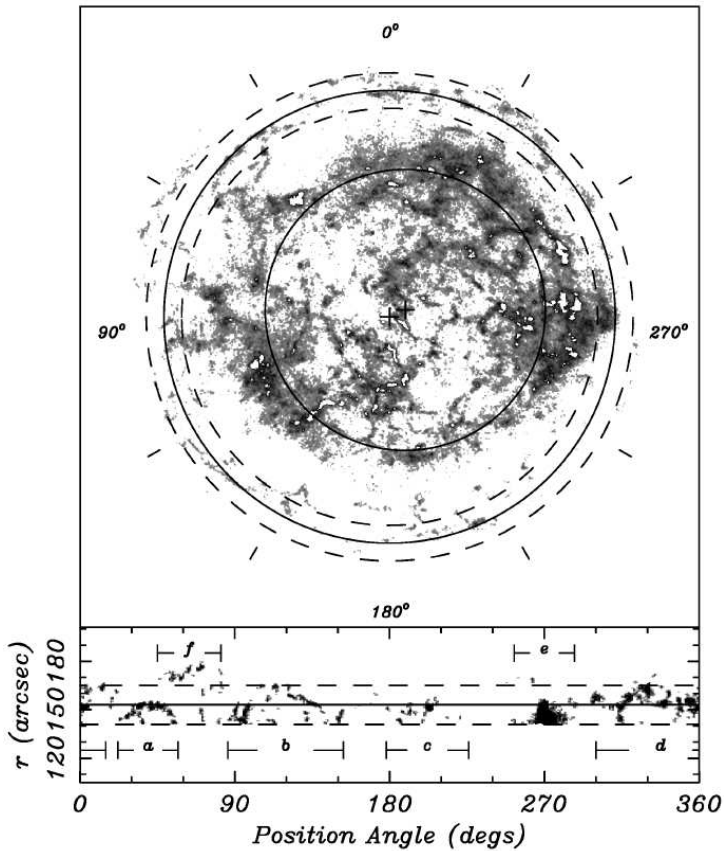,height=4.5in,angle=0,clip=}
}
\caption{ Localization of the forward shocks in \casa. {\it Top panel}
-- the $4-6$ keV high-energy X-ray continuum map with the contrast set
to highlight the previously unresolved fine wisps at the outer edge of
the remnant. We argue that these wisps are the signature of the
forward shock front. The outer circles denote the average distance and
range of the wisps ($153 ^{\prime\prime} \pm 12^{\prime\prime}$).
The inner solid circle shows the mean location of the reverse shock. The
centers for the two circles are indicated by the crosses. {\it Bottom
panel} -- to isolate the wisps, we plot the edge of the remnant in
polar coordinates.  The three lines correspond to the circles in the
top panel. There are $\sim 4$ coherent wisps associated with the forward
show, labled $a-d$ in the bottom panel. 
\label{fig2}}
\end{figure}

\begin{figure} 
\centerline{
\psfig{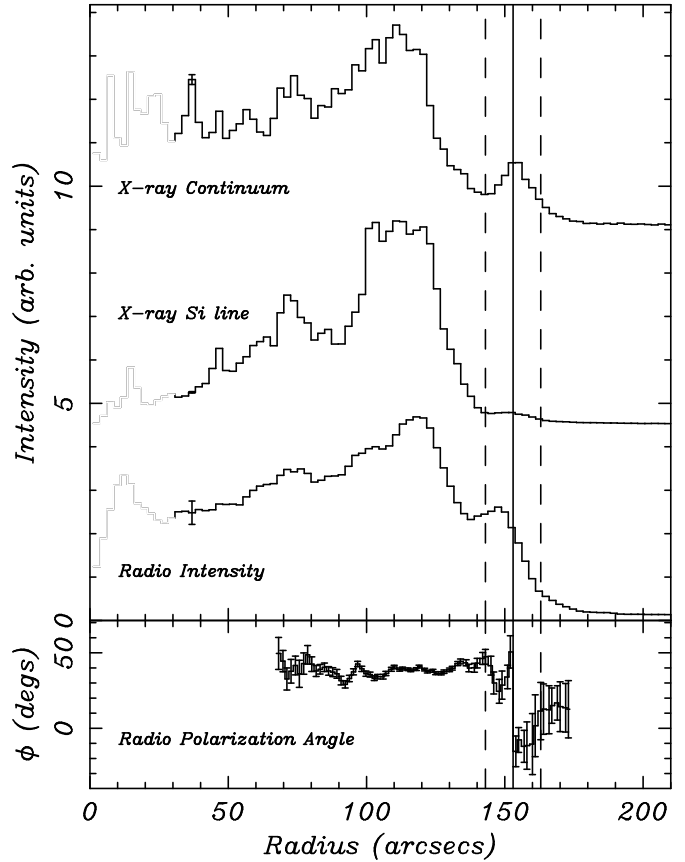}
}
\caption{The behavior of the X-ray and radio radial intensity profiles
of \casa\ near the edge of the remnant.  These profiles are derived
from the images presented in Figure 1 averaged over the northwestern
quadrant of the remnant between $-5$ and $-60$ degs. They are centered
on the wisps (see text) and normalized by area. {\it Top panel} -- A
clear enhancement is seen in the X-ray continuum emission at
$153^{\prime\prime}$ (solid line) around $\pm 12^{\prime\prime}$
(dashed lines) but is nearly absent in the Si emission.
The radio profile shows a sharp rise at the location of the X-ray
edge enhancement. A 1-$\sigma$ error bar is shown; the error decreases
rapidly as a function of radius.  {\it Bottom panel} -- the radio
polarization angle (Epoch 1994) for the above region. The large jump in
angle is found at the X-ray continuum peak. Taken together, the above
profiles suggest that the X-ray wisps found at the edge of the remnant
make up the forward shock front. The 1-$\sigma$ error bar is shown.
\label{fig3}}
\end{figure}

\begin{figure}
\centerline{
\psfig{figure=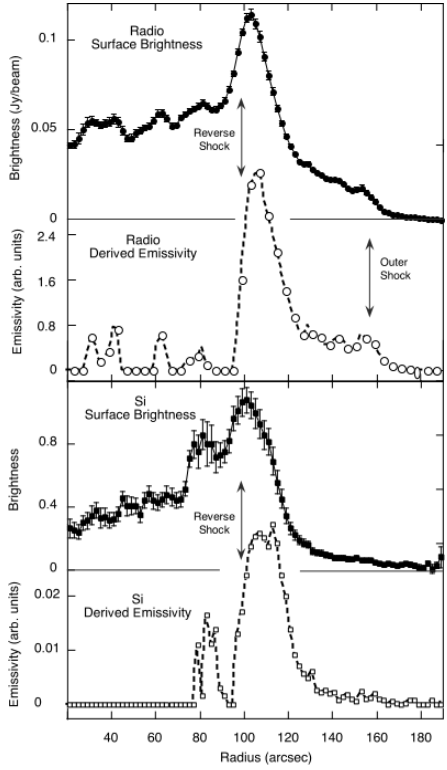,height=4.0in,angle=0,clip=}
}
\caption{ Normalized surface brightness and derived emissivity radial
profiles for radio and Si maps, averaged over azimuths $100-250$
degs.  The derived emissivities are not reliable below $\sim
75\arcsec$.}
\label{fig4}
\end{figure}

\begin{figure}
\psfig{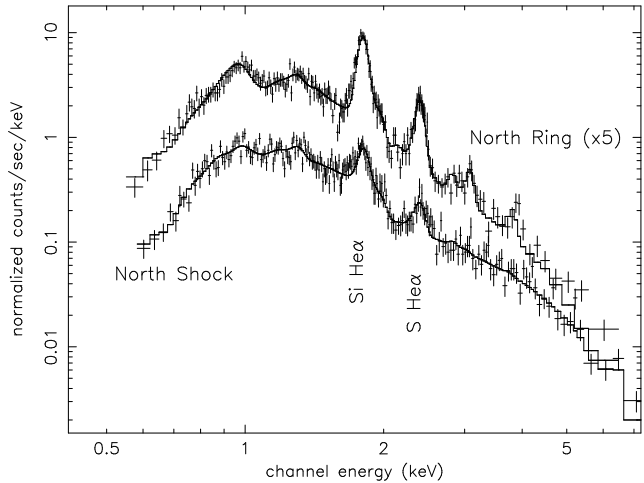}
\caption{\Chandra\ ACIS-S spectra of typical regions in the bright ring (top) and
outer rim (bottom) of \casa.  Equivalent widths of line features
associated with nucleosynthesis products are much larger in the ring spectra,
which is offset ($\times 5$) for clarity.  The solid lines represent 
fits to the spectra, with ranges of parameters as discussed in the text.
\label{fig5}}
\end{figure}

\end{document}